# Motivations and locational factors of FDI in CIS countries: Empirical evidence from South Korean FDI in Kazakhstan, Russia, and Uzbekistan

**Han-Sol Lee**
Department of Marketing,
Peoples' Friendship
University of Russia,
Russia
Email: 1042185141@pfur.ru

**Sergey U. Chernikov**
Department of Marketing,
Peoples' Friendship
University of Russia,
Russia
Email: chernikov-syu@rudn.ru

**Szabolcs Nagy**
Department of Marketing and
Tourism Institute,
University of Miskolc,
Hungary
Email:
nagy.szabolcs@uni-miskolc.hu

Considering the growing significance of Eurasian economic ties because of South Korea's New Northern Policy and Russia's New Eastern Policy, this study investigates the motivations and locational factors of South Korean foreign direct investment (FDI) in three countries in the Commonwealth of Independent States (CIS: Kazakhstan, Russia, and Uzbekistan) by employing panel analysis (pooled ordinary least squares [OLS], fixed effects, random effects) using data from 1993 to 2017. The results show the positive and significant coefficients of GDP, resource endowments, and inflation. Unlike conventional South Korean outward FDI, labour-seeking is not defined as a primary purpose. Exchange rates, political rights, and civil liberties are identified as insignificant. The authors conclude that South Korean FDI in Kazakhstan, Russia, and Uzbekistan is associated with market-seeking (particularly in Kazakhstan and Russia) and natural resource-seeking, especially the former. From a policy perspective, our empirical evidence suggests that these countries' host governments could implement mechanisms to facilitate the movement of goods across regions and countries to increase the attractiveness of small local markets. The South Korean government could develop financial support and risk sharing programmes to enhance natural resource-seeking investments and mutual exchange programmes to overcome the 'red syndrome' complex in South Korean society.

**Introduction**

Over the decades of modern economic development, numerous academic studies have shown the positive impacts of foreign direct investment (FDI) on both host and investing countries (Kindleberger 1969, Buckley–Casson 1976, Hymer 1976, Antonescu 2015, Davletshin et al. 2015, Iwasaki–Suganuma 2015, Duarte et al. 2017, Alhendi et al. 2021). Much of this research demonstrates that multinational enterprises (MNEs) can benefit from production efficiencies, competitive advantages, and access to local production factors through FDI. A high level of FDI inflows has also often contributed to host countries' economic growth by transferring new technologies, increasing productivity, upgrading production facilities, and integrating global economies.

These effects are especially crucial for economies such as South Korea, as the competitiveness of its MNEs is highly dependent on using the locational advantages of other nations to overcome high domestic production costs and a saturated internal market. Competition and the limitations of home markets push South Korean MNEs to participate in globalization with a specific focus on foreign investments. This pattern started in 1968 with the first foreign investment by KODECO in the Indonesian mining sector, and since then South Korea has become the ninth highest investing country (UNCTAD 2020). As shown in Figure 1, excluding two sluggish periods during the Asian and global financial crises, South Korean outward FDI has surged year on year, surpassing 10 billion USD in 2006, 20 billion USD in 2007, 30 billion USD in 2013, 40 billion USD in 2017, 50 billion USD in 2018, and 60 billion USD in 2019.

However, this noticeable development is accompanied by a significant skew in foreign investment destinations. South Korean outward FDI is mostly concentrated on just a few economies: the United States (24%), Cayman Islands (13%), China (9%), Vietnam (7%), and Singapore (5%). This raised little concern in the last decade, as the South Korean economy grew by 3% on average in 2014–2019 and its per capita income surpassed 30,000 USD in 2018. However, in the middle of global economic turbulence, high dependence on a few countries makes the South Korean economy fragile to their decisions and problems (e.g. the 2018 US–China trade war). This suggests that South Korea should hastily change its policy to diversify foreign partnerships to have either a new growth opportunity or a more stable position during the global economic downturn. In this context, the Moon Jae-in government declared the New Northern Policy to expand partnerships with the original 12 Commonwealth of Independent States (CIS)[1] countries.

---

[1] In 2017, the Moon Jae-in government introduced two new foreign policies, namely, the New Southern Policy and New Northern Policy. The New Southern Policy aims to strengthen partnerships with the Association of Southeast Asian Nations (ASEAN) and India. Mongolia and the three northern provinces of China (Liaoning, Jilin, and Heilongjiang) are targets of the New Northern Policy alongside the 12 CIS countries.

Figure 1

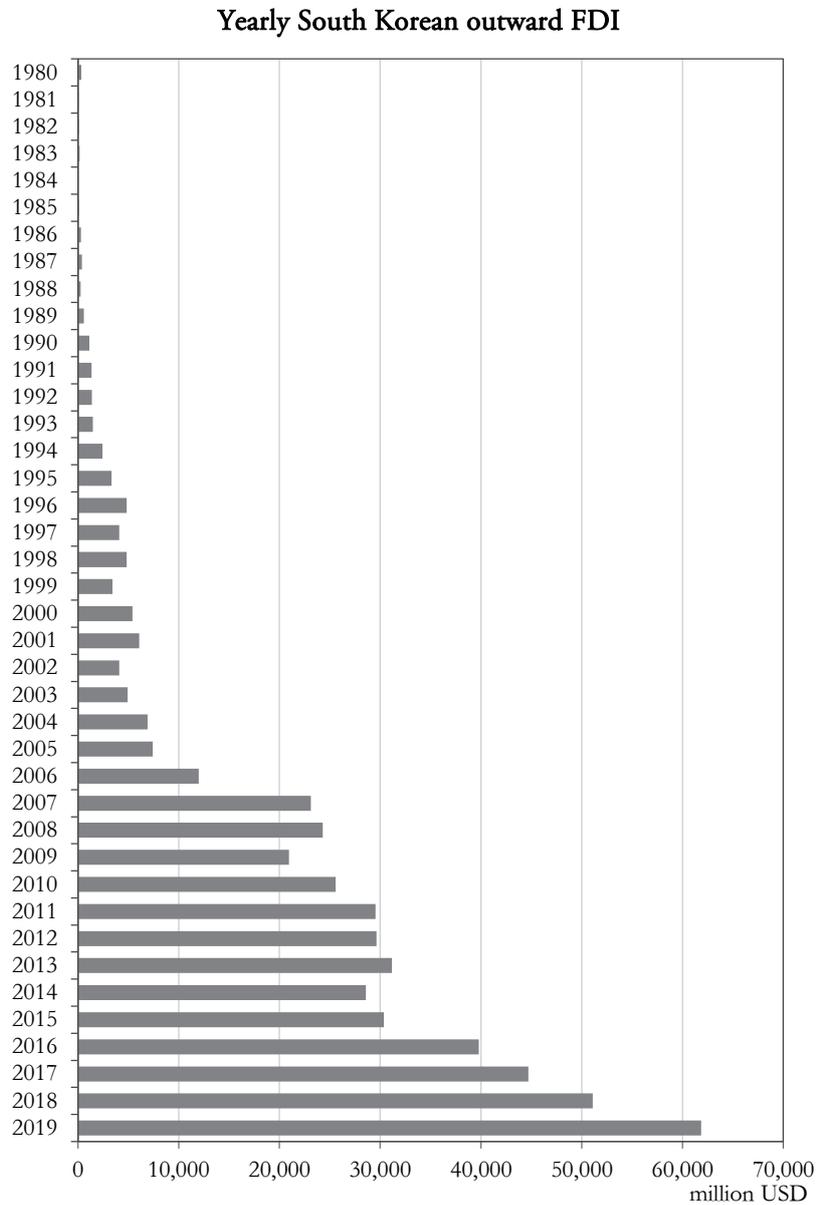

Yearly South Korean outward FDI

*Note:* The value in 1980 is cumulative from 1968 to 1980.
*Source:* The Export-Import Bank of Korea (2020).

CIS transition countries have largely opened their economies since the 1990s to encourage foreign investors through various investment-friendly measures. In 2019, 3.6% of the world's FDI was directed towards transition economies. In particular, Russia was the 15th largest FDI recipient country in the world. As the leading economy of CIS countries, Russia has consistently aimed to build new partnerships with Asia-Pacific countries to overcome the economic stagnation caused by Western sanctions, falling oil prices and national currency (ruble) values, and its energy export-based industrial structure. Since 2012, Russia has enforced the New Eastern Policy and implemented practical policy mechanisms to develop the Russian Far East and reduce dependency on the European economy by attracting foreign investments from neighbouring East Asian countries (mainly China, South Korea, and Japan). This indicates that Russia needs South Korean investments more than at any other point in time. Activating South Korean investments in the largest economy among CIS countries and the Eurasian Economic Union may facilitate business expansion to other CIS countries.

Therefore, considering these economic and political circumstances, enhancing Eurasian economic ties between South Korea and CIS countries has become a crucial agenda. Given the foregoing, defining the motivations and locational factors of South Korean FDI in CIS countries would help increase the effectiveness of further policy implementations by the South Korean and host CIS governments. Moreover, this research contributes to the literature by expanding existing studies of South Korean outward FDI by exploring an uninvestigated case.

For the empirical test, (pooled) ordinary least squares (OLS), fixed effects, and random effects regression analyses are carried out based on data from 1993 to 2017. Among the 12 CIS countries, we select Kazakhstan, Russia, and Uzbekistan, which received South Korean FDI consistently during the study period and have the relevant data available with the necessary quality and history.

The remainder of the paper is organized as follows: dynamics of South Korean FDI in the 12 CIS countries, motivations of South Korea as an FDI investing country, locational factors of CIS countries as FDI hosts, data and hypothesis development, empirical results and discussion of the regression analyses. Finally, we present conclusions and policy implications.

## Dynamics of South Korean FDI in the 12 CIS countries

As shown in Figure 2, South Korean FDI in CIS countries began in 1990 (in Russia, 19 million USD) and expanded to Kazakhstan (0.5 million USD) and Uzbekistan (15 million USD) in 1991 and 1993, respectively. Investment reached 1.3 billion USD in 2008 (accounting for 5.3% of all South Korean outward FDI), but decreased thereafter.

However, despite growing political importance, South Korean outward FDI in CIS countries is still modest: during 1990–2019, only 1.7% of all South Korean FDI was directed to those 12 countries on average and only Kazakhstan, Russia, and Uzbekistan have consistently received South Korean FDI since 1993. South Korean investment has not yet reached Moldova, and Armenia, Belarus, and Turkmenistan have only received South Korean FDI once, in 2017, 2014, and 2015, respectively. Azerbaijan, Georgia, Tajikistan, and Ukraine have also failed to receive consistent South Korean FDI.

Figure 2

### South Korean outward FDI in the 12 CIS countries

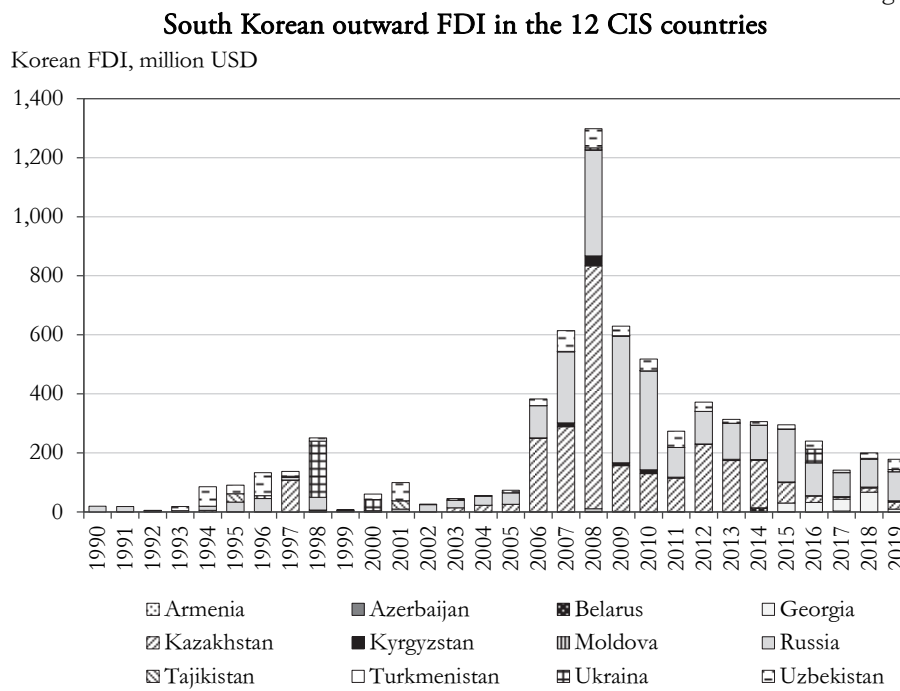

*Source:* The Export-Import Bank of Korea (2020).

We further compare South Korean FDI in Kazakhstan, Russia, and Uzbekistan with that in other BRICs countries (Brazil, Russia, India, and China) and Vietnam (the country that plays a leading role in the New Southern Policy as Russia does in the New Northern Policy) from 2017, the year in which both South Korean policies were introduced.

Table 1

### Comparison of South Korean FDI

| Country | 2017 | | 2018 | | 2019 | |
|---|---|---|---|---|---|---|
| | million USD | % | million USD | % | million USD | % |
| Kazakhstan | 7 | 0.02 | 15 | 0.03 | 24 | 0.04 |
| Russia | 82 | 0.18 | 95 | 0.18 | 99 | 0.15 |
| Uzbekistan | 8 | 0.02 | 18 | 0.04 | 35 | 0.05 |
| Vietnam | 1,987 | 4.43 | 3,342 | 6.50 | 4,585 | 7.13 |
| Brazil | 460 | 1.02 | 394 | 0.77 | 222 | 0.34 |
| India | 516 | 1.15 | 1,072 | 2.09 | 453 | 0.70 |
| China | 3,213 | 7.16 | 4,803 | 9.35 | 5,809 | 9.03 |

*Source:* The Export-Import Bank of Korea (2020).

As shown in Table 1, only modest South Korean FDI was directed to Kazakhstan, Russia, and Uzbekistan (0.03%, 0.17%, and 0.04% of all South Korean outward FDI on average, respectively) for 2017–2019. At the same time, an average of 6.02% of South Korean FDI was directed to Vietnam over the same period. This implies that the New Northern Policy has been implemented less smoothly than the New Southern Policy. Brazil and India, which have worse business conditions (in terms of the World Bank's ease of doing business ranking), received more South Korean FDI than Russia. In 2019, total South Korean FDI in Russia was 99 million USD, while that in Brazil and India was 222 million USD and 453 million USD, respectively. Moreover, South Korean FDI in Brazil showed a similar pattern to that in Russia, as 82% of FDI was concentrated on the manufacturing industry. However, larger amounts of FDI were directed to Brazil than Russia despite the farther geographical distance.

This modest South Korean FDI in Russia compared with Vietnam, Brazil, India, and China, despite advantageous locational factors (market size, business conditions, and geographical distance), can be partially explained by an extant social bias. As Korea has been divided for 70 years because of ideological reasons, a notable social pathology (named the 'red syndrome' complex) has appeared in social and political sectors (Kim 2017). Hence, the low level of South Korean FDI in Russia could be partially attributable to country image stereotyping at various levels of South Korean society. FDI decision-making has a strong hierarchical component in all countries, but especially so in Asia. Considering the average age of major South Korean companies, their image of Russia predominantly goes back to the Soviet Union era when the USSR gave its full support to North Korea. This foundation is enhanced by media coverage on Russia's business and social environments, which focuses on corruption, autocracy, and criminality or elaborates on the information provided by western media. While broader research on the image of Russia among top CEOs in South Korea is clearly needed, 'red syndrome' bias certainly plays a notable role in decisions on FDI direction.

Interestingly, China and Vietnam are only slightly targeted by the 'red syndrome' complex compared with CIS countries. While both adopt a communist ideology, the effects are far less evident, as they share a common Asian culture with South Korea and are not directly connected to the social mindset in North Korea. China accepts the communist ideology for purely economic reasons, while Vietnam is generally considered by South Koreans to have a rather weak position in the international economy and politics. These considerations together with the obvious economic benefits for South Korea using their locational factors (China's market size and low-cost labour in Vietnam) have a clear effect on the amount of FDI.

## Motivations of South Korea as an FDI investing country

Several empirical studies have analysed the motivations of South Korean outward FDI since the 2000s as the government policy has shifted towards encouraging South Korean companies to invest overseas. Most previous studies have employed panel data analysis using a standalone state or country groups. To review those empirical studies, we classify the motivations of South Korean outward FDI following Dunning's (1993) four typologies: market-seeking, resource-seeking, efficiency-seeking, and asset-seeking.

A market-seeking purpose of South Korean outward FDI is consistently found in studies. Lee's (2010) study, based on panel data on South Korean manufacturing firms' FDI for 1980–1996, confirmed the difference in South Korean FDI motivations in developing and developed countries. According to Lee's (2010) results, South Korean outward FDI in developing countries aims to relocate production factors for efficiency-seeking, while that in developed countries is mainly associated with expanding the market. Ha and Lee (2011) asserted that the key purpose of South Korean FDI changed from promoting exports to entering local markets based on a data analysis of 1990–2009. They demonstrated that a host country's gross domestic product (GDP) and export volumes attract South Korean FDI. In the same year, Kim (2011) investigated South Korean FDI determinants in China during 1992–2008 by employing a fixed effects regression. The results confirmed the market-seeking motive (i.e. the positive impact of market size).

Later research has found market-seeking and efficiency-seeking to be the motivating factors of South Korean outward FDI. The results of Gu's (2013) study of the determinants of South Korean FDI in BRICs demonstrated, employing a panel regression, significant results, as they defined goals related to the market size (market-seeking), labour force (labour-seeking), and local currency value (efficiency-seeking). Oh and Mah's (2017) study of the patterns of South Korean outward FDI in Vietnam demonstrated market-seeking and efficiency-seeking motives such as policy changes, a highly educated and low-cost labour force, economic growth, a developed infrastructure, and tax incentives. Seong and Jung (2017) explored South Korean FDI determinants in Trans-Pacific Partnership countries based on the gravity model. They found market expansion as the key objective of South Korean outward FDI, as variables related to production efficiency (i.e. average tariff rate, GDP per capita, remoteness from South Korea) were defined as significant. Similarly, Kim and Yu (2019) investigated the regional distribution of South Korean FDI in 20 Chinese provinces during 2009–2015. The results indicated that South Korean firms tend to invest in Chinese provinces with large markets (market-seeking) and high-quality/low-cost labour (labour efficiency-seeking). Further, they demonstrated that market-seeking goals outweighs labour efficiency-seeking. In addition, some studies have asserted that South Korean outward FDI has an asset-seeking aim. Rugman and Oh (2008) demonstrated that large South Korean MNEs invest in R&D-intensive industries like large western MNEs. Distinct investment features between South Korean conglomerates and small and medium-sized enterprises (SMEs) were also found in Lee's (2006) study. His findings indicated that large South Korean companies invest in high value-added activities for asset-seeking, while cost efficiencies mainly drive foreign investment by South Korean SMEs.

Park and Yang (2011) explored the motivations of South Korean outward FDI based on data from 1980–2000 and identified common and distinguishable purposes before and after the Korean financial crisis. Before the financial crisis, South Korean FDI was associated with market-seeking and asset-seeking, but subsequently switched to seeking efficiency as well. The results also indicated that the importance of economic distance (measured by inter-trade volume) and institutional factors (investment openness) increased in the post-crisis period. In the same vein, Yang and Wang (2017) demonstrated that the purpose of South Korean investment changed from resource-seeking to knowledge-seeking in their study of South Korean FDI in 26 Chinese regions during 2004–2014.

From this literature review, we can draw the following two conclusions. First, market expansion motivates South Korean FDI regardless of the host country and company size. Second, South Korean FDI shows distinct motivations according to the host country's economic status. Investment in a developing country aims to increase production and transaction efficiency, while that in a developed country by South Korean conglomerates aims to obtain strategic assets. However, we should not overlook that South Korea has achieved unprecedented economic growth, taking it from a developing to a developed country. This implies that South Korean FDI does not follow the typical pattern of other developed countries.

South Korean FDI has been more active in less competitive, low capital concentrated, and low product differentiated industries (Kim 1998). This implies that its FDI partially aims to discover less competitive overseas markets and escape the competitive domestic market. In an early study, Dunning (1981, p. 27) categorized ownership advantages into the following three types (as cited in Eden and Dai 2010):
- *"Type 1: advantages that do not arise from multinationality but are advantages that any firm may have over another producing in the same location";*
- *"Type 2: advantages from being part of a multi-plant enterprise";*
- *"Type 3: advantages that come specifically from multinationality."*

Type 1 and Type 2 ownership advantages still exist in the home country. However, Type 3 exists when a company enters a foreign market using a host country's locational factors during internalization. In this sense, a large amount of South Korean companies' FDI lacks home-based ownership advantages but creates ownership advantages by efficiently managing and strategically using the locational factors of the host country. For this reason, South Korean FDI considers the locational factors of a host country to be related to market expansion and cost reduction.

**Locational factors of CIS countries as FDI hosts**

As noted previously, South Korean outward FDI is highly determined by the locational factors of a host country when companies aim to create ownership advantages during internalization. This study further investigates the locational factors of CIS countries as FDI hosts. Owing to the limited availability of FDI data on CIS countries from the 1990s, more studies have been devoted to inward FDI in Russia than in CIS countries (Table 1).

A plethora of studies have identified the significance of market and production factors. In an early study, Ledyaeva (2009) investigated the spatial relationships of FDI inflows in Russia before and after the 1998 financial crisis based on data from 1995 to 2005. The results found that large markets, large cities, high-quality infrastructure (seaports), significant natural resources (oil and gas), closeness to Europe, and lack of political and legislative risks are positive and significant determinants of FDI. The studies by Kayam et al. (2013) and Gonchar and Marek (2013) in Russia also identified market size and natural resource endowments as important factors.

Similar results were presented in other studies of CIS countries. Azam (2010) investigated the key FDI determinants of Armenia, Kyrgyz Republic, and Turkmenistan for 1991–2009, finding a positive influence of market size and official development assistance on FDI and a negative influence of inflation. Kudina and Jakubiak (2012) surveyed 120 enterprises to determine the FDI factors of four CIS countries (Ukraine, Moldova, Georgia, and Kyrgyzstan) before the global financial crisis. The research found that market is a dominant factor behind FDI, while firms are also motivated to use production factors (including natural resources and labour). Akhmetzaki and Mukhamediyev (2017) explored the impacts of Eurasian integration on FDI inflows for 1993–2015 in Eurasian Economic Union member states (Kazakhstan, Belarus, the Russian Federation, Armenia, Kyrgyzstan) and five other countries (Azerbaijan, Tajikistan, Moldova, Georgia, Ukraine). The researchers found that GDP, infrastructure development, and secondary education enrolment positively affect FDI inflows. Conversely, the customs union has a negative impact on FDI inflows.

Governmental policy towards an open economy has been considered to be significant in some work. Yukhanaev et al. (2014) explored regional FDI inflows in Russia during 2005–2011 and defined market size (gross regional product per capita), trade openness, and government policy (special economic zones) as critical determinants. Mariev et al. (2016) investigated the distribution of FDI in 78 Russian regions and 179 investing countries for 2006–2013. The results showed that the GDP of home economies, per capita gross regional product in host regions, openness, innovative capacity, unemployment rate, and distance (from investors to Moscow and from Moscow to regions) are significant.

On the contrary, the impact of factors related to institutional quality has shown mixed results. Cuervo-Cazurra's (2008) research on FDI inflows in transition economies including CIS countries concluded that corruption negatively influences FDI. However, the negative impacts of corruption on FDI differ by subtype: widely known pervasive corruption deters FDI, while unknown arbitrary corruption, perceived as one of the uncertainties of running a business in transition economies, hampers FDI to a lesser degree. Kenisarin and Andrews-Speed (2008) found the positive and significant effects of governance, economic freedom, and corruption perception on FDI to former Soviet Union countries. Kudina and Jakubiak (2012) also defined political instability, volatile economic environments, ambiguous legal systems, and corruption as barriers to FDI.

Some studies have contradicted the conventional view that political and legal risks adversely affect FDI inflows. According to Ledyaeva (2009), political and legislative risks were negative and significant determinants of FDI in Russian regions during the post-financial crisis period (1999–2002). However, neither risk type was identified as statistically significant during 2003–2005, as investors became convinced about Russia's stability. Later, Ledyaeva et al. (2013) examined the effects of corruption and democratization on FDI in Russian regions for 1996–2007. The results showed that less corrupt and more democratic countries are inclined to invest in less corrupt and democratic Russian regions, while foreign investors from corrupt and autocratic countries invest in Russian regions with high corruption and autocracy by collaborating with local politicians. Moreover, in Ulzii-Ochir's (2019) study of the determinants of eight landlocked developing countries in Central Asia for a 20-year period (1996–2016), political instability, corporate tax, and inflation are defined as drivers of FDI inflows, as they create opportunities in countries with a weak and inefficient decision-making process.

In conclusion, foreign investors in CIS countries are mainly attracted by market factors (e.g. market size, per capita growth rate, large cities). Although some studies have defined natural resource factors as insignificant, several authors have shown their positive and significant relation to FDI. In addition to markets and natural resources, a few studies have defined such factors as infrastructure, openness, and taxes as significant. However, the relation between FDI in Russia and institutional quality, defined, for instance, by corruption level, political stability, and legislative risks, shows mixed results. This indicates that the significance of economic and natural resource factors takes precedence over institutional and political factors in Russia.

Table 2

### Summary of the locational factors of CIS countries as FDI hosts

| Study | Countries | Significant locational factors |
|---|---|---|
| Cuervo-Cazurra (2008) | Transition economies including CIS countries | Pervasive corruption deters FDI, while arbitrary corruption does so to a lesser degree. |
| Kenisarin–Andrews-Speed (2008) | Former Soviet Union states | Governance (+), economic freedom (+), corruption perception (+) |
| Ledyaeva (2009) | Russia | Market size (+), large cities (+), infrastructure (+), natural resources (+), distance from Europe (–), political and legislative risks (–) |
| Azam (2010) | Armenia, Kyrgyz Republic, and Turkmenistan | Market size (+), official development assistance (+ in Kyrgyz Republic, Turkmenistan / – in Armenia), inflation rate (–) |
| Kudina–Jakubiak (2012) | 120 enterprises in Ukraine, Moldova, Georgia, and Kyrgyzstan | Market (+), natural resources (+), labour (+), political instability (–), economic volatility (–), ambiguous legal system (–), corruption (–) |
| Ledyaeva et al. (2013) | Russia | FDI from less corrupt and democratic countries flows in less corrupt and democratic regions. However, FDI from corrupt and autocratic countries flows in corrupt and autocratic regions. |
| Kayam et al. (2013) | Russia | Market size (+), natural resources (+), average regional wage (+), infrastructure (+), educational level (+) |
| Gonchar–Marek (2013) | Russia | Market size (+), natural resources (+) |
| Yukhanaev et al. (2014) | Russia | Market size (+), trade openness (+), special economic zones (+) |
| Mariev et al. (2016) | Russia | Market size (+), per capita growth rate (+), openness (+), innovation (+), unemployment rate (–), distance from investors to Moscow and from Moscow to regions (–, the latter does not apply in the remote Far Eastern regions) |
| Akhmetzaki–Mukhamediyev (2017) | Five Eurasian Economic Union members countries | Market size (+), infrastructure (+), secondary education enrolment (+), Eurasian economic integration (–) |
| Ulzii-Ochir (2019) | Eight landlocked developing countries in Central Asia | GDP per capita (+), openness (+), infrastructure (+), political instability (+), corporate tax (+), inflation (+) |

**Data and hypothesis development**

For the regression analysis, we construct balanced panel data on 75 observations from 1993–2017 for Kazakhstan, Russia, and Uzbekistan. For the dependent variable, a natural logarithm of FDI inflows (current million USD) from South Korea to country $i$ in year $t$ is used. The model specification is as follows:

$$LnFDI_{it} = \beta_0 + \beta_1 LnGDP_{it} + \beta_2 LnGGDP_{it} + \beta_3 LnFREE_{it} + \beta_4 LnRESOU_{it} + \beta_5 LnINFLA_{it} + \beta_6 NER_{it} + \varepsilon_{it}$$

(i) $LnGDP_{it}$, the natural logarithm of the GDP of country $i$ in year $t$ (current billion USD), is used as a proxy for market size. Conventionally, larger markets attract more FDI inflows in transition economies (Ledyaeva 2009, Azam 2010, Kudina–Jakubiak 2012, Kayam et al. 2013, Gonchar–Marek 2013, Yukhanaev et al. 2014, Mariev et al. 2016, Akhmetzaki– Mukhamediyev 2017). South Korean outward FDI consistently shows a strong market-seeking purpose (Lee 2010, Ha–Lee. 2011, Kim 2011, Park–Yang 2011, Gu 2013, Oh–Mah 2017, Seong–Jung 2017, Kim–Yu 2019). For descriptive statistics on the variables, see Table A1 in Appendix.

    H0: $LnGDP_{it}$ is positively associated with $LnFDI_{it}$.

(ii) $LnGDPP_{it}$, the natural logarithm of subtracting the GDP per capita of country $i$ from South Korean GDP per capita (current USD) in year $t$, is used as a proxy for the difference in wages from South Korea. During 1993–2017, South Korean GDP per capita was always higher than that of the three CIS countries. Multiple studies have demonstrated that South Korean outward FDI aims at efficiency-seeking (Lee 2006, Lee 2010, Park–Yang 2011, Gu 2013, Oh–Mah 2017, Seong–Jung 2017, Kim–Yu. 2019). Investments tend to flow to countries with low per capita GDP, where labour costs are cheaper than in the domestic market. A larger gap in GDP per capita from that of South Korea to a CIS country indicates higher labour-cost efficiency for South Korean investors in the CIS market.

    H1: $LnGGDP_{it}$ is positively associated with $LnFDI_{it}$.

(iii) $LnFREE_{it}$, the natural logarithm of the sum of political rights ratings and civil liberties ratings of country $i$ in year $t$, is used as a proxy for institutional quality. The quality of institutions does not always lead to FDI inflows in CIS countries (Cuervo-Cazurra 2008, Kenisarin–Andrews-Speed 2008, Ledyaeva 2009, Kudina–Jakubiak 2012, Ledyaeva et al. 2013, Ulzii-Ochir 2019).

    **H2: The coefficient sign of $LnFREE_{it}$ is uncertain.**

(iv) $LnRESOU_{it}$, the natural logarithm of total natural resource rents (% of GDP) of country $i$ in year $t$, is used as a proxy for natural resource endowments. The richness of natural resources is a critical locational advantage for attracting FDI in resource-rich CIS countries (Ledyaeva 2009, Kudina–Jakubiak 2012, Kayam et al. 2013, Gonchar–Marek 2013). As discussed earlier, owing to strong efficiency-seeking, South Korean FDI tends to exploit locational advantages fully.

H3: $LnRESOU_{it}$ is positively associated with $LnFDI_{it}$.

(v) Ln$INFLA_{it}$, the natural logarithm of the inflation rate (annual percentage change in average consumer prices) of country $i$ in year $t$, is used as a proxy for economic stability. High inflation increases macroeconomic instability and investment risks.

H4: $LnINFLA_{it}$ is negatively associated with $LnFDI_{it}$.

(vi) $NER_{it}$, the nominal exchange rate (local currency to USD) of country $i$ in year $t$, is used as a proxy for the purchasing power of investing countries. We convert the original data (USD to local currency) to obtain stationary datasets. An appreciation of host countries' currency value decreases the purchasing power of investing countries. These production cost factors significantly determine efficiency-seeking South Korean FDI.

H5: $NER_{it}$ is negatively associated with $LnFDI_{it}$.

(viii) $\beta_0$ stands for the constant and $\varepsilon_{it}$ represents the error term over time.

## Empirical results and discussion

Table 3 shows the Pearson correlation coefficients between the explanatory variables. LnFREE and LnGGDP (0.61) as well as LnINFLA and NER (0.53) show moderate positive correlations. LnINFLA is negatively and moderately correlated with LnRESOU (–0.50). These variables may cause multicollinearity in a linear function.

Table 3

### Pearson correlations of the explanatory variables, 1993–2017

| Variable | LnGDP | LnGGDP | LnFREE | LnINFLA | LnRESOU | NER |
|---|---|---|---|---|---|---|
| LnGDP | 1.00 | | | | | |
| LnGGDP | 0.01 | 1.00 | | | | |
| LnFREE | –0.39 | 0.61 | 1.00 | | | |
| LnINFLA | –0.38 | –0.44 | –0.27 | 1.00 | | |
| LnRESOU | 0.04 | 0.46 | 0.32 | –0.50 | 1.00 | |
| NER | –0.17 | –0.30 | –0.20 | 0.53 | –0.26 | 1.00 |

To clarify the multicollinearity issue, we carry out a variance inflation factor (VIF) test in the linear function. As no variable has a VIF above 10 in our model, multicollinearity is not considered as an issue in the estimation (Table A2 and A3 in Appendix).

We build the econometric model based on 75 observations of Kazakhstan, Russia, and Uzbekistan from 1993–2017. Table 4 presents the results of the panel analysis. Pooled OLS, fixed effects, and random effects (using the nerlove estimator) are conducted to find the best-fitting model. To compare pooled OLS with fixed effects, we conduct an F-test. As the p-value of the F-test is 0 < 0.05, we choose fixed effects over pooled OLS. To compare fixed effects with random effects, we carry out a Hausman test, finding that the null hypothesis (the random effects model is preferable to the fixed effects model) is accepted at the 5% significance level.

Table 4

### The results of the panel data analysis, 1993–2017

| Independent variable | Type 1 Kazakhstan, Russia, Uzbekistan | | | Type 2 Kazakhstan, Russia | Type 3 Kazakhstan, Uzbekistan |
|---|---|---|---|---|---|
| | (1) Pooled OLS | (2) FE | (3) RE | (4) RE | (5) RE |
| Constant | –14.51** (5.84) | – | –0.94 (6.80) | –8.36 (7.80) | 10.11 (8.01) |
| LnGDP | 0.92*** (0.14) | 2.14*** (0.28) | 2.05*** (0.27) | 1.96*** (0.30) | 2.40*** (0.35) |
| LnGGDP | 0.06 (0.72) | –1.47** (0.72) | –1.35* (0.71) | –0.21 (0.86) | –2.41*** (0.93) |
| LnRESOU | 1.02*** (0.38) | 1.13*** (0.34) | 1.11*** (0.34) | 0.85** (0.43) | 1.71*** (0.40) |
| LnINFLA | 0.32* (0.19) | 0.59*** (0.18) | 0.57*** (0.18) | 0.33* (0.18) | 0.76*** (0.19) |
| LnFREE | 3.71** (1.47) | 0.82 (1.63) | 1.10 (1.61) | | |
| NER | 0.19 (0.62) | –0.01 (0.55) | 0.01 (0.55) | | |
| Pesaran CD (p-value) | 0.19 | 0.46 | 0.45 | 0.55 | 0.75 |
| No. of obs. | 75 | 75 | 75 | 50 | 50 |
| Effect | None | Country | Country | Country | Country |
| $R^2$ | 0.48 | 0.58 | 0.57 | 0.68 | 0.61 |
| Adj. $R^2$ | 0.43 | 0.53 | 0.53 | 0.66 | 0.58 |

Test statistics:
A. (1) Pooled OLS vs. (2) FE: F-test [F=11.26, p-value = 0].
B. (2) FE vs. (3) RE: Hausman Test [chisq = 1.51, p-value = 0.96].

*Note:* Standard errors are given in brackets; the coefficients marked with ***, **, and * are significant at the 1, 5, and 10% levels, respectively.

The results of the RE model for all three CIS countries (model (3)), which shows the best predictability, confirm the significance of LnGDP at the 1% level (support of H0). This indicates that market size is a key driving force of FDI. Further, the LnGGDP coefficient is negative at the 10% significance level (rejection of H1). The gap in per capita income from that of South Korea represents the wages and purchasing power of the local economy. If the labour efficiency-seeking (market-seeking) motivation is strong, the coefficient sign would be positive (negative). In this regard, it can be postulated that market-seeking effects overwhelm labour efficiency-seeking effects for South Korean FDI to CIS countries. LnRESOU is statistically significant at the 1% level (support of H3), confirming that resource endowments motivate FDI.

However, the LnINFLA coefficient is significant and positive at the 1% level, which does not support H4. This implies that economic instability does not negatively impact FDI. According to Ulzii-Ochir's (2019) research on landlocked developing countries in Central Asia, a high inflation rate and having a weak decision-making process do not always hamper FDI because investors aim to seize opportunities during economic turbulence. Another possible explanation is that inflation in CIS countries signifies the opening of their economies in transition periods. Therefore, inflation may somehow have been translated as a market opportunity.

On the contrary, the results also reveal some statistically insignificant variables. The NER coefficient is insignificant and positive. FDI is not associated with efficiency-seeking derived from the local currency value. In addition, the LnFREE coefficient is insignificant and positive. It is likely that market-seeking behaviour is sufficiently predominant to offset any institutional constraints (Jadhav 2012).

In addition, models (4) and (5) are derived from the Type 1 models. As the number of observations decreases from 75 to 50, we rebuild these models with fewer independent variables. In model (4), based on panel data on Kazakhstan and Russia, the LnGDP, LnRESOU, and LnINFLA coefficients are significant and positive at the 1%, 5%, and 10% levels, respectively. Meanwhile, the results of model (5) demonstrate the significant coefficients of LnGDP, LnGGDP, LnRESOU, and LnINFLA. These results also confirm the market-seeking and resource-seeking motivations.

We derive the time-series data of South Korean FDI in Kazakhstan, Russia, and Uzbekistan composed of the four explanatory variables proven to be significant in model (3) to identify the most significant variable in each country.

As shown in Table 5, in Russia, LnGDP has a positive coefficient at the 1% significance level. On the contrary, the other explanatory variables are statistically insignificant in this model. This suggests that market expansion is the foremost motivation of South Korean FDI in Russia, whose domestic markets are relatively more attractive than those in the other CIS countries. In Kazakhstan and Uzbekistan, LnGDP, LnINFLA, and LnRESOU have positive and significant coefficients, indicating that natural resource-seeking is also significant alongside market-seeking. Further, the tendency to use economic turbulence as a business opportunity seems predominant in Kazakhstan and Uzbekistan, but it is stronger in Uzbekistan in terms of the level of significance and coefficient size. While, in Kazakhstan, where market attractiveness is comparable to that in Russia in terms of GDP per capita, market size is the primary factor attracting South Korean FDI.

Considering that LnGDP is an important factor when using reduced variables and observations, it can be postulated that market-seeking is the most predominant motivation of South Korean FDI in CIS countries.

Table 5

### The results of the OLS analysis, 1993–2017

| Independent variable | Kazakhstan | Russia | Uzbekistan |
|---|---|---|---|
| Constant | –2.10 (15.07) | –3.06 (7.43) | 5.39 (13.26) |
| LnGDP | 2.09*** (0.50) | 1.81*** (0.34) | 2.13** (0.90) |
| LnGGDP | –1.01 (1.78) | –0.42 (0.76) | –1.78 (1.73) |
| LnINFLA | 0.49* (0.28) | 0.001 (0.23) | 1.03*** (0.30) |
| LnRESOU | 1.66** (0.69) | –0.48 (0.47) | 1.58** (0.63) |
| Breusch–Pagan–Godfrey test (P-value) | 0.60 | 0.68 | 0.35 |
| Breusch–Godfrey test (P-value) | 0.52 | 0.59 | 0.69 |
| No. of obs. | 25 | 25 | 25 |
| $R^2$ | 0.73 | 0.75 | 0.41 |
| Adj. $R^2$ | 0.67 | 0.70 | 0.30 |

*Note:* Standard errors are given in brackets; the coefficients marked with ***, **, and * are significant at the 1, 5, and 10% levels, respectively.

## Conclusions

As the New Northern Policy was declared as one of the 100 state affairs in 2017, CIS countries' roles in the future economic growth of South Korea (as diversifying foreign partnerships is highly skewed towards the United States and China) have recently become pivotal. However, South Korean outward FDI in CIS countries remains insignificant because of inefficient policies. Further, although economic cooperation between South Korea and CIS countries (particularly Russia because of the New Eastern Policy) is becoming increasingly important, no previous studies have explored South Korean FDI in CIS nations. In this respect, this study contributes to the literature by analysing the motivations and locational factors of South Korean FDI in three CIS countries, namely, Kazakhstan, Russia, and Uzbekistan, based on data on 1993–2017.

From the results of the econometric analysis, we can draw the following conclusions. First, we found that South Korean FDI's primary goal in Kazakhstan,

Russia, and Uzbekistan is market-seeking and natural resource-seeking, particularly the former. South Korean FDI in those three countries is positively facilitated by GDP and inflation, and negatively by GDP per capita gap (South Korea – a local country). This implies that not economic stability but markets and new business opportunities are the predominant factors attracting South Korean FDI. The significance of GDP is predominant, particularly in Russia and Kazakhstan, whose domestic markets are large and attractive relative to those in other CIS countries. While natural resource endowments also significantly attract South Korean FDI, the impact is predominant in Kazakhstan and Uzbekistan, whereas factors related to institutional quality (political rights and civil liberties) do not influence South Korean FDI.

Second, South Korean FDI's motivation in Kazakhstan, Russia, and Uzbekistan is not associated with efficiency-seeking (i.e. using cheap labour costs and local currency values), which contradicts conventional findings on the motivations of South Korean FDI. Therefore, reducing production costs is not the primary purpose of South Korean FDI in Kazakhstan, Russia, and Uzbekistan.

Our empirical evidence suggests future policy implications that could enhance South Korean FDI in CIS countries for both the South Korean and the host countries' governments. First, CIS countries' governments should improve the attractiveness of their local markets. Excluding some major cities (e.g. Moscow, Saint Petersburg, Astana, and Almaty), many cities in CIS countries have small purchasing teams of regional buyers. Therefore, investment aiming for market expansion like South Korean FDI may target not only a region of their business entry but other neighbouring regions and countries (exporting to third-party countries). The spatial relation is a critical factor when deciding where to locate firms (Cao 2021). Hence, CIS countries' governments should construct mechanisms to facilitate the movement of goods between regions and countries by simplifying customs clearance processes and improving distribution infrastructure. For this, we recommend that they could improve efficiencies in the customs clearance system (e.g. unification of tariff rates and HS codes), reduce the number of documents and inspections, and automate the customs clearance process. We also suggest establishing an inter-governmental committee to develop joint research and investment on road infrastructure to improve inland distribution channels. To this end, host governments should ease regulations and lower barriers to entry in transportation industries.

Second, the South Korean government should develop financial programmes to support natural resource-seeking investments that require considerable funds and include high risks compared with market-seeking investments. A project requiring high capital (e.g. in the energy sector) might be reliant on governmental subsidies (Lados et al. 2020). Further, as natural resource investments are closely related to national energy and political security, the South Korean government could facilitate natural resource-seeking investments by expanding the reliable suppliers of natural resources and deepening economic cooperation. To do this, it could implement various financial support packages (e.g. establish inter-governmental investment funds and joint-venture funds with multilateral development banks) and risk-sharing programmes (e.g. minimum revenue guarantee, minimum cost support, and credit security policies).

Lastly, the implementation of policies to increase mutual exchanges in the private sector could overcome the 'red syndrome' complex in South Korean society and promote investments in the long term. To do this, South Korean and CIS companies could develop mutual internship programmes to spur continuous exchanges. In addition, the South Korean government could consider using new media platforms to show daily life in CIS countries and provide information on their economics, politics, history, and travel.

## Appendix

Table A1

### Descriptive statistics, 1993–2017

| Variable | Source | N | Mean | S.D. |
|---|---|---|---|---|
| LnFDI | The Export-Import Bank of Korea | 75 | 3.0564 | 2.0535 |
| LnGDP | IMF | 75 | 4.6177 | 1.7103 |
| LnGGDP | IMF | 75 | 9.5060 | 0.3565 |
| LnFREE | Freedom House | 75 | 2.4223 | 0.1860 |
| LnINFLA | IMF | 75 | 2.9604 | 1.4216 |
| LnRESOU | World Bank | 75 | 2.6874 | 0.5811 |
| NER | Federal Reserve Bank of St. Louis | 75 | 0.0908 | 0.3434 |

Table A2

### The results of the VIF test (panel analysis), 1993–2017

| LnGDP | LnGGDP | LnFREE | LnRESOU | LnINFLA | NER |
|---|---|---|---|---|---|
| 1.80 | 2.02 | 2.32 | 1.50 | 2.30 | 1.41 |

Table A3

### The results of the VIF test (OLS analysis), 1993–2017

| Country | LnGDP | LnGGDP | LnRESOU | LnINFLA |
|---|---|---|---|---|
| Kazakhstan | 3.13 | 2.35 | 1.61 | 1.90 |
| Russia | 2.57 | 1.54 | 1.43 | 3.02 |
| Uzbekistan | 7.17 | 7.46 | 2.76 | 2.57 |